\documentclass{article}

\usepackage{PRIMEarxiv}

\usepackage[utf8]{inputenc} 
\usepackage[T1]{fontenc}    
\usepackage{url}            
\usepackage{xurl}
\usepackage{xcolor}
\usepackage[colorlinks = true,
            linkcolor = purple,
            urlcolor  = cyan,
            citecolor = cyan,
            anchorcolor = black]{hyperref}
\usepackage{booktabs}       
\usepackage{amsfonts}       
\usepackage{nicefrac}       
\usepackage{microtype}      
\usepackage{graphicx}       
\usepackage{float}
\usepackage{placeins}

\usepackage{amsmath}                
\usepackage{subcaption}             
\usepackage{tabularx}               
\usepackage{array}                  
\usepackage{enumitem}               

\graphicspath{{figures/}}

\usepackage{tikz}
\definecolor{gsblue}{HTML}{4285F4}
\DeclareRobustCommand{\gsicon}{%
	\begin{tikzpicture}[baseline=-0.35em]
	\draw[gsblue, fill=gsblue] (0,0) circle [radius=0.16];
	\node[white] at (0,0) {\fontfamily{phv}\selectfont\bfseries\tiny G};
	\end{tikzpicture}%
}

\newcommand{\scholarauthorA}{vU6oBhwAAAAJ}
\newcommand{\scholarauthorB}{R98vFWoAAAAJ}
\newcommand{\scholarA}{\hspace{0.25em}\href{https://scholar.google.com/citations?user=\scholarauthorA}{\gsicon}}
\newcommand{\scholarB}{\hspace{0.25em}\href{https://scholar.google.com/citations?user=\scholarauthorB}{\gsicon}}

\def\blind{0}

\title{A Data-Driven Analysis for Engineering Conferences: the Institute of Industrial and Systems Engineering (IISE) Annual Conference Proceedings (2002--2025)
\thanks{\textit{\underline{Citation}}:
\textbf{H.S. Bank and C. Eaton. A Data-Driven Analysis for Engineering Conferences: the Institute of Industrial and Systems Engineering (IISE) Annual Conference Proceedings (2002--2025). In Proceedings of the 2026 IISE Annual Conference, Arlington, TX, USA, May 16--19, 2026.}}
}

\if0\blind
\author{
  H. Sinan Bank\thanks{Corresponding author: sinan.bank@colostate.edu}\scholarA \\
  Department of Systems Engineering \\
  Colorado State University \\
  Fort Collins, CO 80523, USA \\
  \texttt{sinan.bank@colostate.edu} \\
  \And
  Casey E. Eaton\scholarB \\
  Department of Industrial and Systems Engineering \\
  Auburn University \\
  Auburn, AL 36849, USA \\
  \texttt{cee0051@auburn.edu} \\
}
\fi

\if1\blind
\author{
  Author information is purposely removed for blind review
}
\fi

\begin{document}
\maketitle

\begin{abstract}
Charting the intellectual evolution of a scientific discipline is crucial for identifying its core contributions, challenges, and future directions. The IISE Annual Conference proceedings offer a rich longitudinal archive of the Industrial and Systems Engineering (ISE) community's development, but the sheer volume of scholarship produced over two decades makes a holistic analysis difficult. Traditional reviews often fail to capture the full scale of thematic shifts and complex collaboration networks that define the community's growth. This paper presents a computational analysis of IISE proceedings from 2002 to 2025, drawing on an initial dataset of 9,350 titles from ProQuest for thematic analysis and 8,958 titles from Google Scholar for citation analysis, to deliver a cartography of the ISE field's intellectual history. Leveraging Large Language Models (LLMs) for domain-aware classification, Natural Language Processing, and Network Science, our study systematically maps thematic evolution to identify dominant, emerging, and receding research topics. We analyze citation data and co-authorship networks to uncover influential papers and authors, providing critical insights into knowledge diffusion and community structure. Through this comprehensive analysis, we establish a baseline for understanding the trajectory of ISE research and offer valuable insights for researchers, practitioners, and educators. The findings illuminate the field's intellectual assets and provide a data-informed map to guide the future of ISE. To foster reproducibility and further research, the curated dataset used in this study and the results will be made publicly available.
\end{abstract}


\section{Introduction}

Conferences play a critical role in disseminating research, especially in engineering and computer science disciplines where they are a primary venue for publishing original results \cite{kochetkov2021importance}. While conference proceedings are vital for the rapid spread of ideas and account for a substantial 10\% of engineering literature, they tend to be cited less and become outdated sooner than journal articles \cite{lisee2008conference}. Despite these constraints, a comprehensive bibliometric analysis of an engineering field is therefore incomplete without examining the field's conference proceedings, as they capture the emerging topics and community engagement that are often absent from journals.

The central motivation for this work is to create the first large-scale, data-driven intellectual history of the modern Industrial and Systems Engineering (ISE) field by analyzing its flagship conference. At this scale, systematically analyzing over two decades of proceedings presents significant methodological challenges. For instance, the choice of a scholarly database fundamentally shapes the results. Broad-coverage sources (e.g., Google Scholar, etc.) are invaluable for capturing publications missed by curated indexes, but they suffer from significant data quality and reproducibility issues, making large-scale analysis difficult \cite{halevi2017suitability}. In contrast, curated databases (e.g., ProQuest, etc.) offer structured metadata but may have a limited scope and lack the citation links necessary for impact analysis.

To address these challenges, we combine ProQuest's structured metadata with Google Scholar's broader citation coverage, creating a robust multi-source dataset spanning 2002 to 2025. This dual-database approach enables us to validate findings across sources while maximizing both data quality and comprehensiveness. We employ LLM-based classification to track how research priorities have shifted across different ISE domains (e.g., manufacturing, healthcare, logistics) and identify which topics are gaining or losing momentum. Beyond thematic trends, our network analyses reveal the underlying social and intellectual fabric of the community: citation patterns highlight knowledge flows and theoretical foundations, while co-authorship networks illuminate collaborative ecosystems.

\section{Relevant Work}

Our research is grounded in the established practices of bibliometric analysis, leveraging computational techniques to provide a quantitative and objective overview of a scientific field. The importance of such data-driven analysis in ISE is well-established; researchers were identifying ``big data'' as a revolutionary force as early as 2013, with applications ranging from sentiment analysis on social media platforms \cite{hutchison2013big} to transforming business performance and decision-making by moving from intuition to data-driven rigor \cite{mello2014big}. Our work builds on this principle by applying a similar data-centric lens not to a business, but to the research field of ISE itself. This study synthesizes methodologies from two key domains: the application of Natural Language Processing (NLP) for thematic analysis and the use of Network Science to map community structure while maintaining a commitment to reproducible research.

\textbf{Natural Language Processing for Thematic Analysis:} Natural Language Processing (NLP) techniques enable systematic analysis of thematic content in large scholarly corpora, ranging from traditional probabilistic models such as Latent Dirichlet Allocation (LDA) \cite{blei2003latent} to modern transformer-based methods (e.g., BERTopic \cite{grootendorst2022bertopic}). Comparative studies demonstrate BERTopic's better performance, achieving coherence scores of 0.56 compared to LDA or Non-Negative Matrix Factorization (NMF) while requiring less manual refinement \cite{ebrahimi2025latent}. Recent advances show open-source LLMs combined with semantic search can extract domain-specific knowledge from multi-document corpora with over 90\% accuracy \cite{ghali2024empowering}, while keyword-based clustering techniques effectively identify research interconnections and thematic patterns in bibliometric analyses \cite{aruwajoye2024bibliometric}. These diverse computational approaches provide potential assets for extracting thematic insights from scholarly literature.

\textbf{Network Science for Mapping Community Structure:} Beyond thematic content, network science methods enable analysis of research community structure and collaboration patterns. This methodology examines two primary constructs: citation data, which reveal influential papers and the lineage of ideas, and co-authorship networks, which illuminate the social structure of collaboration. Citation data has been used to identify influential researchers and characterize knowledge flows \cite{guo2018network}, while co-authorship analysis can reveal collaboration patterns, the emergence of new researchers, and the internationalization of research communities. These network approaches provide complementary perspectives on the intellectual and social dimensions of scholarly fields.

These methodological pillars form the foundation of our approach. We now detail how their integration enables a comprehensive, longitudinal analysis of the IISE Annual Conference proceedings, producing a systematic cartography of ISE research evolution spanning two decades.

\section{Methodology}

Our analytical approach follows a systematic four-phase workflow adapted from established systematic review methodologies \cite{kitchenham2007guidelines}. Phase 1 (Planning) defined the research scope and protocol, establishing temporal boundaries (2002--2025) and selecting ProQuest and Google Scholar as complementary data sources to balance coverage breadth with metadata quality. Phase 2 (Collection and Selection) implemented structured queries, yielding 9,350 records from ProQuest and 8,958 from Google Scholar, followed by deduplication (using fuzzy title matching with 85\% similarity threshold), quality filtering, and standardization of author names with cross-database matching. Phase 3 (Analysis) applied computational methods to extract insights from three data dimensions: (a) thematic content through LLM-based classification into domains and sub-areas, (b) citation networks from Google Scholar data (focusing on top 200 most cited papers), and (c) co-authorship patterns from both ProQuest and Google Scholar datasets. Phase 4 (Results) synthesizes these perspectives to characterize research evolution, influential contributions, and collaboration structures. This integrated approach enables comprehensive longitudinal analysis while maintaining reproducibility through transparent protocols. We detail the core methodological components below.

\textbf{Data Collection and Preparation:} We adopted a dual-database strategy to address the fundamental tension between data coverage and quality. ProQuest served as the primary source for thematic analysis, providing structured metadata and comprehensive IISE proceedings coverage through queries targeting the ``IISE Annual Conference Proceedings'' source (2002--2025). This extraction yielded 9,350 records including titles, abstracts, authors, keywords, and publication years. Google Scholar complemented this with broader citation coverage, yielding 8,958 records with citation counts and bibliographic metadata. Data preparation followed a three-stage process: (1) deduplication using DOI matching and fuzzy title matching (using rapidfuzz library with token\_sort\_ratio scorer at 85\% similarity threshold) to merge records across databases, (2) quality filtering to remove incomplete records, and (3) standardization of author names (normalized to last name + first initial format) with cross-database matching based on last name and paper co-occurrence (Jaccard similarity). This process ensured a high-quality, validated dataset for computational analysis.

\textbf{Thematic Analysis Through LLM-Based Classification:} To systematically map research themes and their evolution, we employed Large Language Models (LLMs) for domain-aware paper classification. Our approach first extracted a domain knowledge base from IISE Annual Conference's Call for Abstracts (CfA) (one per research domain as provided in PDF), to guide LLM-assisted text extraction for identifying domains and sub-areas. This knowledge base was structured hierarchically, with domains (e.g., ``Data Analytics and Information Systems'', ``Manufacturing and Design'') containing multiple sub-areas (e.g., ``Machine Learning'', ``Computer Vision''). We then classified papers using LLM-based prompts implemented via LangChain, comparing local models (e.g., Llama, Mixtral, DeepSeek, etc.) and cloud-based models (e.g., Gemini). The classification process matched papers to domains first and then sub-areas. Each classification included a primary domain, secondary domain (if applicable), primary sub-area, and confidence level (High/Medium/Low). This approach enabled fine-grained thematic mapping while maintaining interpretability through explicit domain-sub-area relationships. Temporal analysis tracked domain and sub-area prevalence (percentage of papers) over time, identifying dominant themes (consistently high prevalence), emerging themes (increasing prevalence), and declining themes (decreasing prevalence).

\textbf{Network Analysis:} We constructed and analyzed two complementary network types to map the intellectual and social structure of the ISE research community. Citation networks were built from Google Scholar data, focusing on the 200 most cited papers. The network is directed, with edges representing citation relationships (citing paper $\rightarrow$ cited paper). We computed standard network metrics including in-degree (citation counts for influence), out-degree (number of references), PageRank scores (recursive influence), and betweenness centrality (bridging papers). Temporal analysis of citation patterns reveals knowledge flows and identifies periods of consolidation versus diversification.

The curated dataset is archived in Harvard Dataverse \cite{bank2026dataset} and the codebase is shared via a GitHub repository \cite{bank2026codebase} with documentation. A web application for interactive visualization and use of the analysis outcomes is available \cite{bank2026webapp}. The codebase includes Jupyter notebooks for each analysis phase: data preprocessing, citation network analysis, co-authorship network analysis, and LLM-based thematic classification. Processed datasets include author networks (e.g., GraphML format), citation networks, and paper classifications with domain and sub-area assignments. This multi-platform approach ensures that our analysis serves as a reusable resource for the ISE research community and enables independent verification and extension of our findings.

\begin{figure}[H]
    \centering
    \includegraphics[width=\textwidth]{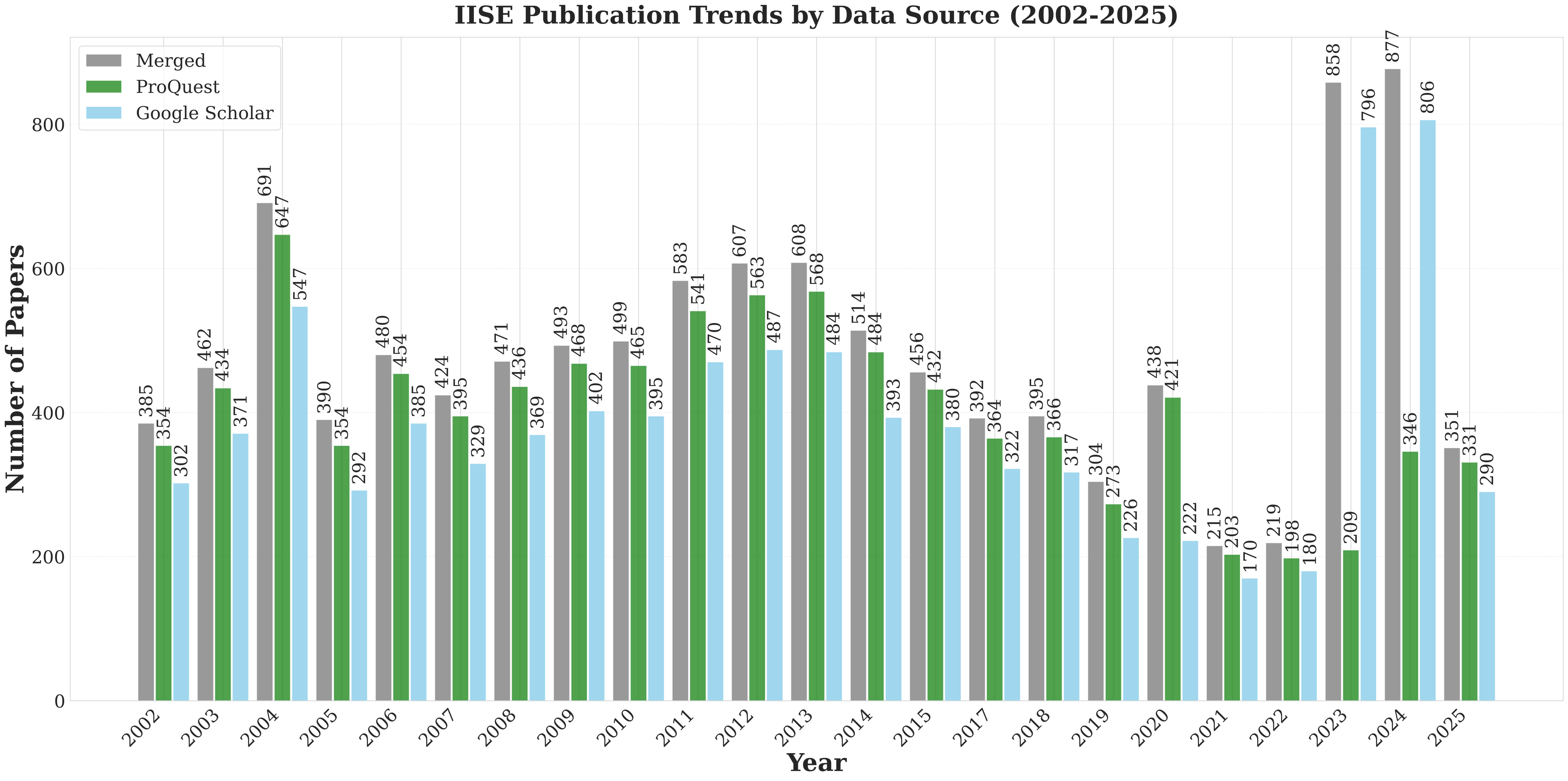}
    \caption{Annual IISE Conference publication counts by data source (2002--2025). 2016 excluded due to data unavailability in ProQuest and Google Scholar.}
    \label{fig:publication_trends}
\end{figure}

\section{Results}

This section presents the descriptive statistics characterizing publication trends, then examine thematic evolution through LLM-based domain and sub-area identification and classification, analyze knowledge diffusion via citation networks, map collaboration structures through co-authorship networks, and conclude with integrated insights connecting these complementary perspectives.

\textbf{Publication Trends and Descriptive Statistics:} Figure \ref{fig:publication_trends} presents the temporal distribution of IISE Annual Conference proceedings publications across the study period. The analysis shows distinct phases of activity rather than linear growth: an initial expansion (2002--2004), a decade of relative stability (2005--2015), a pre-pandemic contraction, and a dramatic post-pandemic resurgence. The average number of presentations per year is approximately 485, with notable peaks in 2004, 2012, and the 2023--2024 period corresponding to expanded conference tracks and the community's robust recovery following the COVID-19 disruptions (2020--2021). Notably, the discrepancy between data sources---where ProQuest captures more records in earlier years while Google Scholar offers broader coverage in specific later periods---validates the necessity of the multi-source aggregation strategy.

\begin{figure}[H]
    \centering
    \includegraphics[width=\textwidth]{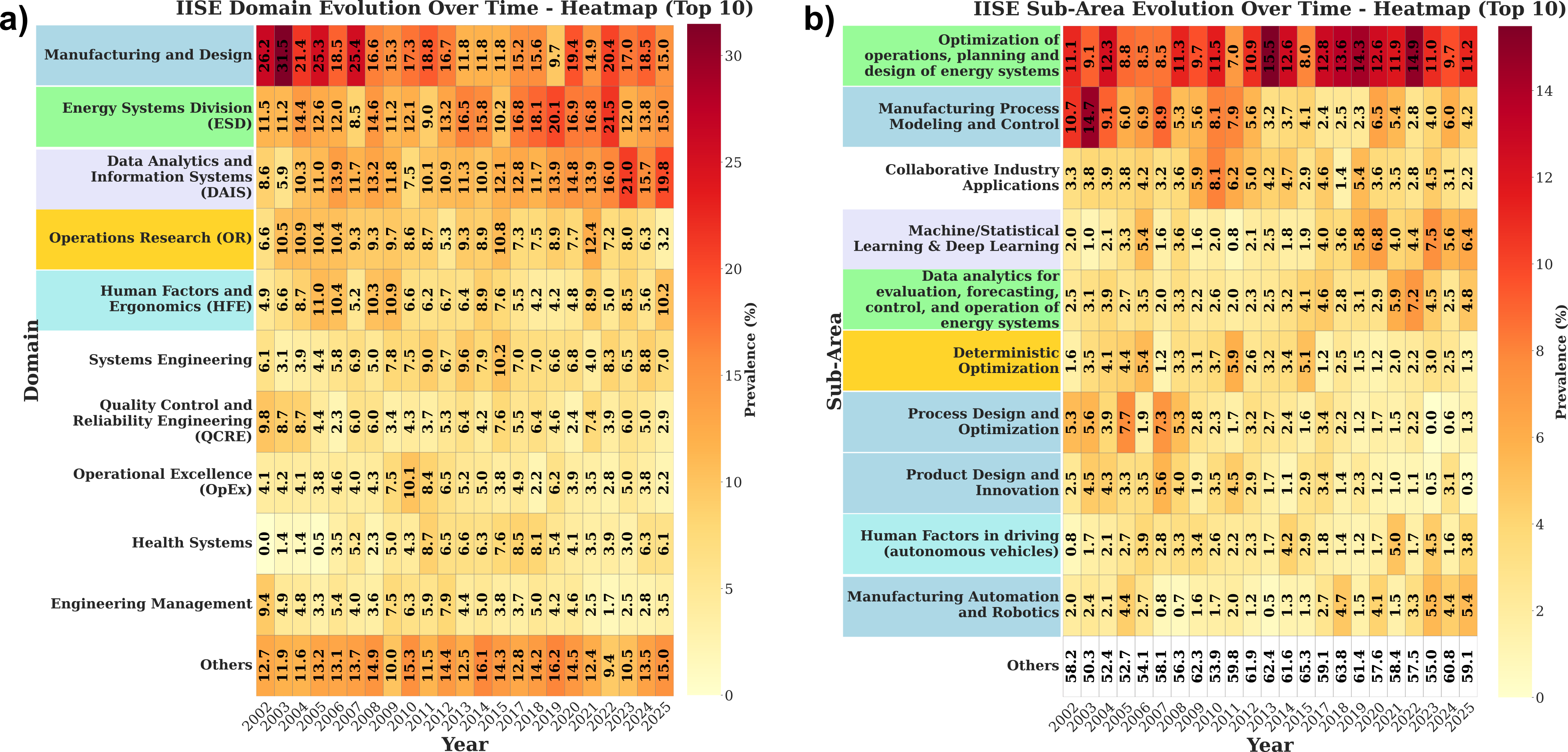}
    \caption{Major research \textbf{a)} domains and \textbf{b)} sub-areas in IISE proceedings (2002--2025) --- Mixtral:8x22b.}
    \label{fig:topic_evolution}
\end{figure}

\textbf{Thematic Evolution:} LLM-based classification of thematic evolution---comparing multiple models as detailed in \cite{bank2026codebase}---identified 16 research domains with 246 sub-areas across the corpus, aligned with the IISE Annual Conference 2026 CfAs as the ground truth. Validation of domain and sub-area identification against this ground truth yielded the following performance metrics: all tested models achieved perfect domain classification (F1-Score: 1.0; Precision: 1.0; Accuracy: 1.0; Recall: 1.0), while Mixtral:8x22b achieved the reasonable sub-area classification performance (Accuracy: 0.808; F1-Score: 0.894; Precision: 0.894; Recall: 0.894) with the fastest total estimated process time (9 hrs 5 mins). Based on these metrics of Mixtral:8x22b, we concluded that the model's understanding of the ISE concept is on par with the other models while completing its prediction in shortest amount of time. Following this validation, we applied domain and sub-area identification and classification to the full publication set. Fig.\ \ref{fig:topic_evolution} shows the temporal evolution of the top 10 domains and sub-areas across all publications, revealing a paradigm shift in the IISE Annual Conference proceedings. The domain-level analysis (Fig.\ \ref{fig:topic_evolution}a) illustrates a transition from traditional physical systems to data-centric and service-oriented methodologies. Manufacturing and Design, while remaining a core pillar, exhibits a gradual relative decline from a peak of 31.1\% in 2003 to 14.8\% in 2025. Conversely, Data Analytics and Information Systems (DAIS) demonstrates a growth, rising from 8.6\% in 2002 to become the dominant domain at 18.9\% in 2025, effectively doubling its footprint. Similarly, Health Systems evolved from a negligible presence (0.0--1.4\% in the early 2000s) to a steady 6.0\% share, highlighting the successful integration of ISE principles into healthcare. Sub-area analysis (Fig.\ \ref{fig:topic_evolution}b) corroborates this digital transformation. The Data Analytics sub-area shows a consistent trajectory (4.5\% to 11.7\%), while Machine Learning \& Deep Learning emerges as a distinct, rapidly growing category in the latter half of the decade, reaching 4.7\% by 2025. In contrast, despite the increase in Manufacturing Automation and Robotics, the traditional branch of modeling areas such as Manufacturing Process Modeling show a downward trend in relative prevalence (dropping from 10.2\% to 3.8\%), suggesting that while these topics remain foundational, the frontier of research innovation has shifted toward computational and data-driven applications.

\begin{figure}[H]
    \centering
    \includegraphics[width=\textwidth]{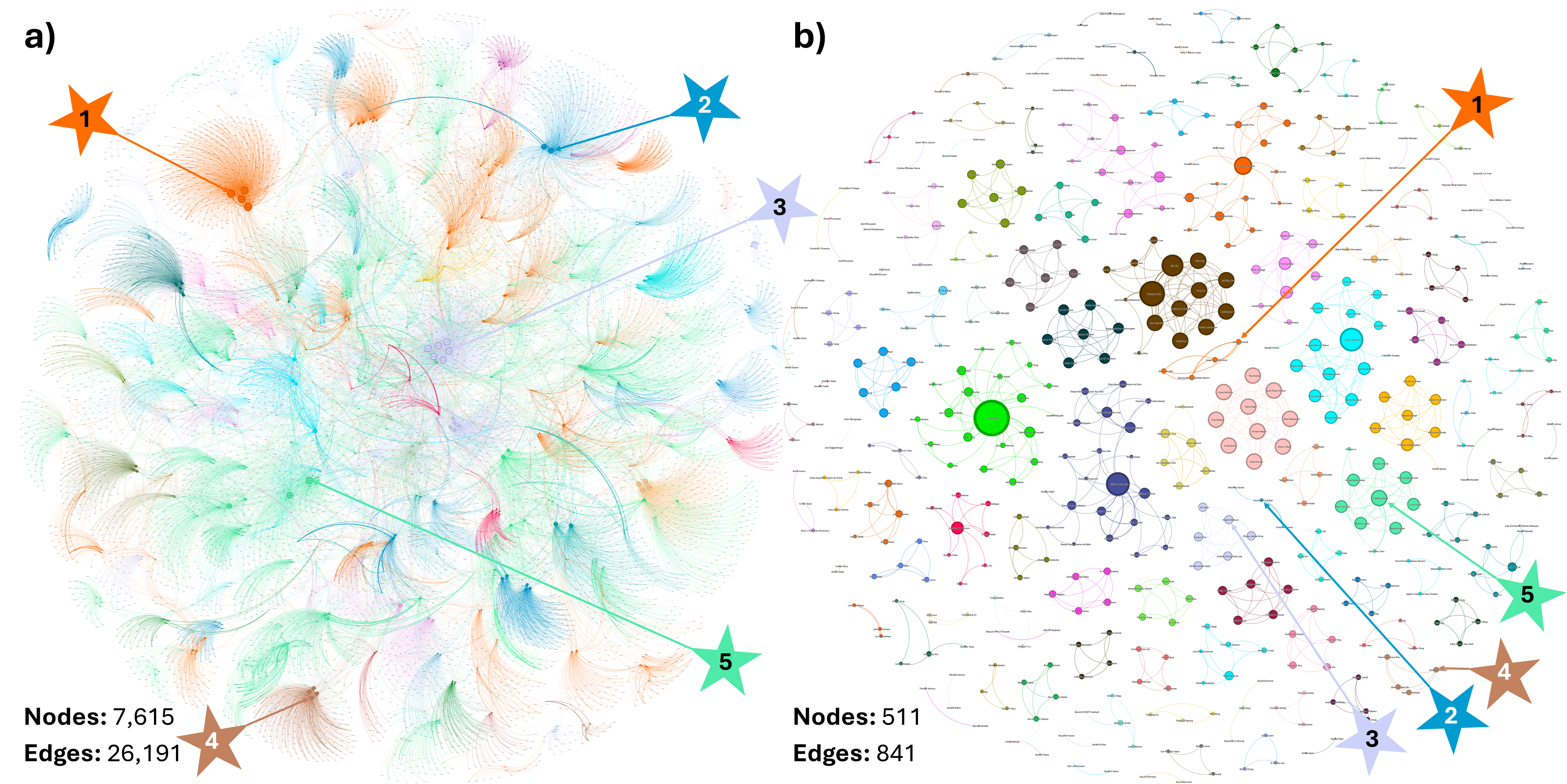}
    \caption{Author-level \textbf{a)} citation and \textbf{b)} co-author networks for top 200 cited papers between 2002 and 2025.}
    \label{fig:citation_coauthor_networks}
\end{figure}

\textbf{Citation and Co-author Network Analysis:} Citation and co-author network analysis (data as of November 8, 2025) identified the most influential papers and authors within the IISE proceedings corpus based on multiple centrality metrics (e.g., eigenvector, betweenness, etc.) as detailed in Table \ref{tab:centrality_metrics} and \cite{bank2026codebase}. Table \ref{tab:influential_papers} lists the top 5 most cited papers. These contributions span diverse areas including health systems, operational excellence, and operation research, with citation counts ranging from 96 to 135. For co-authorship networks, we employed three centrality measures to characterize author influence: degree centrality quantifies the number of direct co-authorship connections, eigenvector centrality identifies authors connected to well-connected collaborators, and betweenness centrality highlights authors who serve as bridges connecting different research communities. Table \ref{tab:centrality_metrics} presents the top 3 authors for each metric from the network of top 200 cited papers. Figure \ref{fig:citation_coauthor_networks} presents the author-level citation network and co-author network, where the first authors from Table \ref{tab:influential_papers} are highlighted with the same ID numbers.

\begin{table}[ht]
    \centering
    \begin{subtable}{\textwidth}
        \centering
        \caption{Top 5 most cited papers in IISE Annual Conference proceedings (2002--2025).}
        \label{tab:influential_papers}
        \begin{tabularx}{\textwidth}{c >{\raggedright\arraybackslash\hsize=1.0\hsize}X >{\raggedright\arraybackslash\hsize=1.15\hsize}X >{\raggedright\arraybackslash\hsize=0.85\hsize}X c c}
        \toprule
        \textbf{\#} & \textbf{Title} & \textbf{Authors} & \textbf{Domain} & \textbf{Year} & \textbf{Cites} \\
        \midrule
        1 & Medical devices... & Jamshidi, Afshin et al. & Health Systems & 2014 & 135 \\
        2 & A3 reports: tool for... & Sobek II, Durward K et al. & Operational Excellence & 2004 & 122 \\
        3 & A robust ensemble-deep... & Maftouni, Maede et al. & Health Systems & 2021 & 101 \\
        4 & Decentralized approaches... & Dolinskaya, Irina S et al. & Operation Research & 2011 & 97 \\
        5 & Breast cancer prediction... & Wang, Haifeng et al. & Health Systems & 2015 & 96 \\
        \bottomrule
        \end{tabularx}
    \end{subtable}

    \vspace{0.3cm}

    \begin{subtable}{\textwidth}
        \centering
        \caption{Top 3 authors by degree and centrality metrics of co-authorship network (2002--2025) from the top 200 cited papers.}
        \label{tab:centrality_metrics}
        \begin{tabularx}{\textwidth}{c >{\raggedright\arraybackslash}X c | c >{\raggedright\arraybackslash}X c | c >{\raggedright\arraybackslash}X c}
        \toprule
        \multicolumn{3}{c|}{\textbf{Degree}} & \multicolumn{3}{c|}{\textbf{Eigenvector Centrality}} & \multicolumn{3}{c}{\textbf{Betweenness Centrality}} \\
        \cmidrule(lr){1-3} \cmidrule(lr){4-6} \cmidrule(lr){7-9}
        \textbf{ID} & \textbf{Author} & \textbf{Value} & \textbf{ID} & \textbf{Author} & \textbf{Value} & \textbf{ID} & \textbf{Author} & \textbf{Value} \\
        \midrule
        D1 & S Desai & 20 & E1 & YJ Son & 0.34 & B1 & RA Wysk & 0.0018 \\
        D2 & YJ Son & 14 & E2 & J Liu & 0.33 & B2 & S Ramakrishnan & 0.0017 \\
        D3 & LF McGinnis and EM Van Aken & 13 & E3 & AM Khaleghi & 0.30 & B3 & G Okudan Kremer & 0.0015 \\
        \bottomrule
        \end{tabularx}
    \end{subtable}
\end{table}

These integrated results provide a comprehensive cartography of ISE research as reflected in IISE Annual Conference proceedings, revealing both the intellectual evolution and the social structure that drives innovation in the field.

\section{Conclusion and Future Work}

A comprehensive understanding of the IISE Annual Conference proceedings reveals the intellectual evolution and collaborative structure of the ISE community over two decades. The findings provide a foundation for evidence-based decision-making regarding future research directions and educational priorities in the field.

While our LLM-based classification approach provides systematic and relatively scalable thematic analysis, several limitations should be acknowledged such as variations across different LLM models, divergence from human expert judgments, etc. We provide some details in the \cite{bank2026codebase} while leaving more of the details for future work.

Given the rapid evolution of computational methods, particularly agentic workflows powered by Large Language Models that can automate systematic literature analysis at unprecedented speed \cite{cao2025automation}, conducting exhaustive manual studies repeatedly becomes impractical. The fast pace of research and technological change suggests that future work should leverage these automated approaches to create ``living analyses'' that continuously update as new proceedings emerge. Our study provides a comprehensive baseline cartography against which such dynamic, agent-driven systems can be validated and compared, establishing the foundation for more sustainable and scalable approaches to monitoring the intellectual evolution of the ISE community. The methodology, data \cite{bank2026dataset}, and codebase \cite{bank2026codebase} presented here can be readily expanded to other conferences and journals to extract more in depth insights (e.g., combining the outcome with the federal funding landscape), with particular potential for application to AIIE-IIE-IISE Transactions (1969--2025), which would provide a complementary longitudinal view of the field's evolution through peer-reviewed journal publications.


\bibliographystyle{unsrt}
\bibliography{IISE2026_DataDrivenAnalysis}

\end{document}